\documentstyle[12pt,graphics,epsfig]{article}

\newcommand{\D}{\displaystyle}
\newcommand{\be}{\begin{equation}}
\newcommand{\ee}{\end{equation}}
\newcommand{\bea}{\begin{eqnarray}}
\newcommand{\eea}{\end{eqnarray}}
\newcommand{\nn}{\nonumber}

\newcommand{\del}{\nabla}
\newcommand{\tU}{\tilde{U}}
\newcommand{\teta}{\tilde{\eta}}
\newcommand{\tg}{\tilde{g}}
\newcommand{\om}{\omega}

\newcommand{\br}{{\bf r}}
\newcommand{\bb}{{\bf b}}
\newcommand{\bR}{{\bf R}}
\newcommand{\bu}{{\bf u}}
\newcommand{\bs}{{\bf s}}
\newcommand{\beff}{{\bf f}}
\newcommand{\rp}{r^{\prime}}

\newcommand{\brp}{{\bf r}^{\prime}}
\newcommand{\omp}{\omega^{\prime}}
\begin{document}
\title{The Wolf effect and the Redshift of Quasars}
\author{Daniel F. V. James,\\
\small{ Theoretical Division T-4, Mail Stop B-268,}\\ 
\small{Los Alamos National Laboratory, Los Alamos, NM 87545, USA} }
\date{\today}
\maketitle
\begin{center}
To be published in {\em Journal of the European Optical Society 
A: Pure and Applied Optics}, \\ Special issue on Physical Optics and 
Coherence Theory in honor of Professor Emil Wolf 
\\(A. T. Friberg and D. F. V. James, editors).
\end{center}

\begin{abstract}
We consider a simple model, based on currently accepted
models for active galactic nuclei,  for a quasi-stellar object 
(QSO or ``quasar'')
and examine the influence that correlation-induced spectral changes
(``The Wolf Effect'') may have upon the redshifts of the optical 
emission lines.\\
LAUR-98-417
\end{abstract}


\section{Introduction}
\setcounter{equation}{0}

The Wolf effect is the name given to several closely related
phenomena in radiation physics dealing with the modification of the
power spectrum of a radiated field due to {\em spatial} fluctuations
of the source of radiation.  The first paper to demonstrate a
connection between spatial source fluctuations (as
characterized by the cross-spectral density function) appeared in
1987 \cite{T:2}.  Since then over 100 papers have appeared on this
and related topics (for a recent review, see ref. \cite{R:7}).  

The discovery of coherence-induced spectral shifts grew out of
work investigating the connection between optical coherence theory
and the empirical laws of radiometry, in particular the 
dependence of radiative transfer on the wavelength of light.  In
1986, Wolf introduced the scaling law \cite{E:1}, which is
a condition upon the spatial coherence properties of a source
under which the radiated field has the same normalized spectrum
as the source; in other words, if the source obeys the scaling
law, the radiation pattern is independent of wavelength.  From this
discovery, it was a comparatively short step to investigate the 
properties of a radiation source which does not obey the scaling law; 
it was found that for such a source there would be a small shift 
of the central frequencies of spectral lines, with shifts towards
lower frequencies predominating \cite{E:2}.  This effect was soon
confirmed experimentally \cite{E:1,E:2,E:4,E:6}.  

Because of the well known analogy between radiation and scattering,
it is not surprising that analogous effects may occur in the
scattering of light.  Such spectral shifts were first investigated
in ref.\cite{S:1} for the case of static (i.e. time invariant) 
scattering media; 
the analysis of the case of time-dependent scatterers followed
soon after \cite{S:2}.  This led to an important realization that
scattering from time-dependent fluctuating random media could
mimic the Doppler effect in many of its important features \cite{S:3}.
An example of such a scattering medium which could
produce Doppler-like shifts was found in 1990: it involved
a highly {\em statistically anisotropic}
\footnote {we will use the term {\em statistically anisotropic}
for such media to avoid confusion with other types of anisotropies,
for example media which respond to different polarizations
of light in different ways} scattering medium in the sense that the
coherence length of the fluctuations of the scatterer was much
longer in one direction than in the others \cite{S:6} (see
also \cite{S:5,S:7,S:11}).  

It was speculated in ref.\cite{E:2} that this effect might 
play a role in the interpretation of the 
spectra of certain extra-galactic astronomical objects such as 
quasars.  Since then other applications, notably remote sensing, 
communications and filtering have since be proposed; However in
this paper we will return to this first application.
In the work on scattering from anisotropic media, it was pointed out
that anisotropies of the type under consideration are compatible
with current models of quasars \cite{S:6}.  Here we will investigate
these points further by proposing a model for a quasar, closely 
related to a model currently favored \cite{peterson:97}, which
predicts a sizable spectral shift due to the Wolf effect. The
rest of the paper is organized as follows: in section 2 we discuss
briefly various types of spectral shift phenomena and their importance
to cosmology.  In section 3 we give a qualitative discussion of
the model quasar we are considering and a mathematical analysis of the
spectrum of the light it radiates.  
Finally we assess the possible implications of the results derived.

\section{Spectral shifts and Cosmology}
In this section we will give a brief and non-exhaustive description of
different physical phenomena which can give rise to shifts in the 
central frequency of a spectral line.  The magnitude of such a shift 
is often measured in terms of the fractional shift or $z$-number, 
defined by the formula
\be
z=\frac{\lambda_{0}^{\prime}-\lambda_{0}}{\lambda_{0}}
\equiv \frac{\omega_{0}-\omega_{0}^{\prime}}{\omega^{\prime}_{0}}, 
\label{zee}
\ee
where $\lambda_{0}$ is the {\em unshifted} central wavelength of the 
spectral line; 
$\lambda_{0}^{\prime}$ is its {\em shifted} central wavelength;
$\omega_{0}$ is the angular frequency corresponding to $\lambda_{0}$ 
and $\omega_{0}^{\prime}$ is the angular frequency corresponding to
 $\lambda^{\prime}_{0}$.  When $z$ is a positive number the spectral
 line has been shifted to a lower frequency (i.e. a redshift) and
 when $z$ is negative the shift is to a higher frequency (a blueshift).
 From eq.(\ref{zee}) it is clear that $z$ must always be in the 
 range $-1<z<\infty$.

Possibly the most famous physical phenomenon which can produce spectral
shifts is the Doppler effect.  When a source of radiation is in motion
with respect to the observer, there is a shift of the wavelength of
spectral lines with respect to the wavelength that would be observed
if the source were at rest.  In this case the $z$-number is given by
the the following formula \cite{doppler:1842,einstein:05}:
\be
z=\frac{1-\beta\cos\theta}{\sqrt{1-\beta^{2}}} -1
\approx-\beta\cos\theta , 
\ee
where $\beta=v/c$, $v$ being the speed of the source relative to the 
observer and
$c$ the speed of light, and $\theta$ is the angle between the
velocity of the source and the line joining the source and
the observer.  The approximate expression is valid when $\beta\ll1$.
For example, if the source is moving directly away from the
observer, so that $\theta=\pi$, the $z$-number has the property
$z>0$, i.e. a redshift is observed, whereas if the source
is heading directly toward the observer (so that $\theta=0$)
then the $z$-number is negative and a blueshift will be observed.  
An important
property of the Doppler effect is that the $z$-number is
{\it independent} of the wavelength of the line in question.
The fact that redshifts with $z$-numbers approximately
independent of wavelength are observed in the spectra of
all galaxies, thereby suggesting that the spectral
shift is due to the Doppler effect associated with
receding motion, is the single most compelling piece of
evidence in favor of the expanding universe.  Hubble's
law, the empirical linear relationship between distances and redshifts 
of close-by
galaxies, confirms the predictions of cosmological
models based on general relativity, and so it has become
widely accepted that the redshift of the radiation emitted by an object is an
indicator of its cosmological distance.  However there is
observational evidence that seems to contradict these assumptions 
\cite{arp:87}.
A well known
example is the close pairing of the galaxy NGC-4319 (with
redshift $z=0.006$) and the quasar Mk-205 ($z=0.07$), 
which, despite having drastically different redshifts
(and therefore being a completely different distances
according to the usually accepted interpretation of
redshifts), have a distinct luminous connection between
them.  If we abandon this conventional interpretation,
we must then ask what phenomena, other
than the recessional motion of an object due to cosmological
expansion, may be the cause of the observed spectral shifts.

Another important type of spectral shift is the gravitational
shift discovered by Einstein \cite{einstein:11}.  This occurs
when there is a difference between the gravitational potential
in the vicinity of the source and the gravitational potential
in the vicinity of the  observer.  In this case the $z$-number is
given by 
\be
z=\frac{\Delta\Phi}{c^{2}}
\ee
where $\Delta\Phi$ is the difference in gravitational potential.
This shift, like the Doppler shift, has a $z$-number which is
independent of wavelength.  However, it can be shown that this
shift is small for quasars.

Another type of shift, which is sometimes suggested as a possible
source of spectral shifts in quasars, is the Compton effect, due to
scattering by free electrons \cite{compton:23a,compton:23b}.  
The scattered radiation has a spectral shift with respect to
the incident radiation given by the formula:
\be
z=\frac{2\hbar\omega_{0}\sin^{2}\left(\Theta/2\right)}{m_{e}c^{2}}
\label{zcomp}
\ee
where $m_{e}$ is the mass of an electron and $\Theta$ is the scattering
angle. However, shifts due to the Compton effect have $z$-numbers
which {\it are} dependent on the wavelength of incident radiation
(i.e. $\omega_{0}$, the angular frequency of the the incident
radiation, appears in eq.\ref{zcomp}), and therefore the Compton 
effect cannot account for all of the large shifts observed in the
spectra of extra-galactic objects.

Other important optical effects which produce spectral shifts are
the closely related scattering phenomena known as Brillouin scattering
\cite{brillouin:21} and Raman scattering \cite{raman:28a,raman:28b}.
Both of these effects produce changes in the wavelength of light
due to interactions with excitations of internal degrees of freedom
of a scattering medium, such as rotational excitations of molecules
or vibrations of a crystal.   However these effects do not produce 
simple spectral shifts: the spectrum of scattered monochromatic
radiation contains various discrete frequencies (a triplet in the
case of Brillouin scattering, multiplets in the case of Raman 
scattering).  

One other hypothesized type of spectral shift should
also be mentioned, namely the redshift due to fundamental
particles having variable
masses \cite{greenberger:74,hoyle:74,arp:90}.  Although the effect
has never been confirmed experimentally, it can be put
on a sound theoretical basis, and it does seem to offer
a possible explanation for many otherwise contradictory 
observations of the universe \cite{arp:87}. 

To these possibilities, we now add the Wolf effect, already discussed
briefly in the introduction.  In the following section we discuss
a model of  a quasar which displays a redshift of its spectral
lines due to this effect.

\section{A model for active galaxies with discordant redshifts}
\setcounter{equation}{0}

Quasars are star-like objects which have many unusual properties.  
They emit a non-thermal continuum spectrum of radiation and emission
lines that are highly redshifted ($z$-numbers of up to 4 or 5 have 
been reported).  Because of this high redshift, they are conventionally
thought to be very distant objects, which therefore can reveal 
important information about the structure of the universe on very
large scales.  However, given their observed intensity,
if this interpretation of their redshifts were correct, then quasars must
be very powerful sources of radiation indeed.  Since the discovery of quasars
over 30 years ago, very elegant theories have evolved to explain their
properties and account for this peculiar intensity of radiation.  The 
possibility that the observed intensity might be the correct indicator
of distance (rather than redshift), and that therefore these
object might have a peculiar non-cosmological redshift is
usually implicitly rejected.

The basic features of currently fashionable models of quasars
\cite{peterson:97}, and 
other types of peculiar extra-galactic object (collectively
known as active galactic nuclei or AGNs), are shown in figure
1.  The most important point is that these objects have axial
rather than spherical symmetry.  At the center is a very dense
object (the `central engine', sometimes characterized as a black 
hole, although no direct confirmation of the predictions of general
relativity have been seen) which, due to gravitational attraction,
sucks in matter from the dust torus, causing it to radiate with
a broad, non-thermal power spectrum.  This radiation excites gaseous
clouds which will emit line radiation.  Usually there are thought
to be two types of clouds, the broad line regions (BLR) and the narrow
line regions (NLR), which emit spectral lines of different widths.
(In our diagram we have lumped the BLR and NLR together).  The beauty 
of this model is that the properties of different types of AGN can be 
explained quite simply by the fact that one is observing the same type 
of object from different directions.  The dust torus blocks out certain
wavelengths in certain directions.  

Our model differs from that usually considered in two important 
ways.  First, the line emitting clouds are considerably closer to the
central engine (or alternatively the dust torus has a considerably 
larger bulge), so that the line radiation is collimated as well as
the continuum radiation.  Secondly we assume the existence of a 
time fluctuating scattering medium, whose properties will be discussed below, which 
causes the line radiation to be scattered into the direction of the
observer.  A schematic version of this model, suitable for 
mathematical analysis, is shown in fig.2

\subsection{spectrum of scattered radiation}
Scattering by a time-dependent medium is described by the following 
inhomogeneous partial differential equation:
\be
\left(\del^{2}-\frac{1}{c^{2}}\frac{\partial^{2}}{\partial t^{2}}\right)
U(\br,t)=-4\pi\eta(\br,t)U(\br,t).
\label{waveqn}
\ee
In this formula, $U(\br,t)$ is the radiation field, treated in the
scalar approximation, and $\eta(\br,t)$ is the time-dependent 
dielectric susceptibility of the scattering medium.  We have
used a simplified model for the medium, in which $\eta(\br,t)$
has only a single time argument; more generally, it should
have two time arguments \cite{wolf:110}, in order to account for
the possibilities of resonant frequencies of the scatterer.

We will analyze the scattering in the
space-frequency domain.  The Fourier transforms \footnote{Since both $U(\br,t)$
and $\eta(\br,t)$ are stationary random processes, generalized 
harmonic analysis  is required.} of $U(\br,t)$  and $\eta(\br,t)$
are defined by the formulas:
\bea
\tU(\br,\omega)&=&\frac{1}{2\pi}\int^{\infty}_{-\infty}U(\br,t)
\exp(i\om t) dt,  \nn \\
\teta(\br,\om)&=&\frac{1}{2\pi}\int^{\infty}_{-\infty}\eta(\br,t)
\exp(i\om t) dt . 
\label{ftdef}
\eea
Substituting these definitions into
eq. (\ref{waveqn}) we obtain the following partial differential equation:
\be
\left(\del^{2}+k^{2}\right)
\tU(\br,\om)=-4\pi\int^{\infty}_{-\infty}\teta(\br,\om-\omp)U(\br,\omp) 
d\omp ,
\label{waveqnft}
\ee
where $k=\om/c$.
This can be solved using the standard Green's function:
\bea
\tU(\br,\om)&=&\tU_{0}(\br,\om)\nn \\
&&+
\int\int d^{3}\rp d\omp \teta(\brp,\om-\omp)
\tU(\brp,\omp) \frac{\exp (ik\left|\br-\brp\right|)}
{\left|\br-\brp\right|},
\eea
where $\tU_{0}(\br,\om)$ is the incident field.
This is a Fredholm integral equation of the second kind.  It can
be solved using the standard Born series.  Retaining the
first two terms only, we obtain the following expression
for the scattered field:
\bea
\tU_{s}(\br,\om)&\equiv&\tU(\br,\om)-\tU_{0}(\br,\om) \nn \\
&\approx&
\int\int d^{3}\rp d\omp \teta(\brp,\om-\omp)
\tU_{0}(\brp,\omp) \frac{\exp(ik\left|\br-\brp\right|)}
{\left|\br-\brp\right|} . 
\eea
In the far zone (i.e. in the limit
$|\br|\gg|\brp|$), the scattered field is given by the
formula:
\bea
\tU_{s}^{(\infty)}(r\bu,\om)
&\approx&\frac{\exp(ikr)}{r}
\int\int d^{3}\rp d\omp \teta(\brp,\om-\omp)
\tU_{0}(\brp,\omp) \exp (-ik \bu\cdot\brp) . \nn \\
\eea
where $\bu$ is the unit vector in the direction of observation (fig.2).
The power spectrum of the far-zone scattered field is therefore:
\bea
S^{(\infty)}(r\bu,\om)&=&\frac{1}{r^{2}}\int\int
d^{3}\rp_{1}d^{3}\rp_{2}\int d\omp
W_{\eta}(\brp_{1},\brp_{2},\om-\omp)\nn \\
&&\times
W_{0}(\brp_{1},\brp_{2},\omp)
\exp[ik\bu\cdot(\brp_{1}-\brp_{2})]
\eea
where we have used the following identities involving the 
cross-spectral density functions for the incident field,
$W_{0}(\brp_{1},\brp_{2},\om)$, and for the dielectric 
fluctuations $W_{\eta}(\brp_{1},\brp_{2},\om)$ and the
spectrum of the scattered field $S^{(\infty)}(r\bu,\om)$:
\bea
\left\langle\tU_{s}^{(\infty)\ast}(r\bu,\om)
\tU_{s}^{(\infty)}(r\bu,\omp)\right\rangle &=& S^{(\infty)}(r\bu,\om)
\delta(\om-\omp) \nn \\
\left\langle\tU_{0}^{\ast}(\brp_{1},\om)\tU_{0}(\brp_{2},\omp)
\right\rangle &=& W_{0}(\brp_{1},\brp_{2},\om)
\delta(\om-\omp) \nn \\
\left\langle\teta^{\ast}(\brp_{1},\om)\teta(\brp_{2},\omp)
\right\rangle &=& W_{\eta}(\brp_{1},\brp_{2},\om)
\delta(\om-\omp) ,
\eea
the chevron brackets denoting an average over ensembles
of {\em both} the field and dielectric fluctuations.

\subsection{The Incident Field}
Assume that the incident field is radiated by a three
dimensional primary
source distribution $Q(\br,t)$.  Further, assume that the
scattering region is in the far zone of the source.
The cross-spectral density of the field is then given by
\bea
W(r_{1}\bs_{1},r_{2}\bs_{2},\om)&=&\frac{\exp[ik(r_{2}-r_{1})]}
{r_{1}r_{2}}\int\int d^{3}r_{1}^{\prime\prime}d^{3}r_{2}^{\prime\prime}
W_{Q}(\br_{1}^{\prime\prime}, \br_{2}^{\prime\prime}, \om)\nn \\
&&\times
\exp[ik(\bs_{1}\cdot\br_{1}^{\prime\prime}-
\bs_{2}\cdot\br_{2}^{\prime\prime})] ,
\label{ringo}
\eea
where the cross-spectral density function $W_{Q}(\brp_{1},\brp_{2},\om)$
of the source distribution is defined by 
\be
\left\langle \tilde{Q}^{*}(\brp_{1},\om)
\tilde{Q}(\brp_{2},\omp) \right\rangle
= W_{Q}(\brp_{1}, \brp_{2}, \om)\delta(\om-\omp)
\ee
where $\tilde{Q}(\brp,\om)$ is the generalized Fourier transform of
the source distribution $Q(\br,t)$.

It will be convenient for the later
development to express this result in terms of
 ``sum-and-difference'' coordinates, as follows:
\bea
{\mathcal W}(R\bs,\bb,\om)&\approx& \frac{\exp[ik\bs\cdot\bb]}
{R^{2}}
\int\int d^{3}R^{\prime\prime}d^{3}r^{\prime\prime}
{\mathcal W}_{Q}(\bR^{\prime\prime}, \br^{\prime\prime}, \om) \nn \\
&&\times
\exp\left[-ik\left(\bs\cdot\br^{\prime\prime}+
\frac{\bb_{\perp}\cdot\bR^{\prime\prime}}{R}\right)\right]
\label{george}
\eea
where 
\be
{\mathcal W}(\bR,\bb,\omp)\equiv 
W\left(\bR-\frac{1}{2}\bb,\bR+\frac{1}{2}\bb,\om\right)
\ee
and
\bea
\bR\equiv R\bs=\frac{r_{1}\bs_{1}+r_{2}\bs_{2}}{2};\,\,\,\,\,\,\,
\bb&=&r_{2}\bs_{2}-r_{1}\bs_{1} \nn \\
\bR^{\prime\prime}=\frac{\br_{1}^{\prime\prime}+\br_{2}^{\prime\prime}}{2}
;\,\,\,\,\,\,\,
\br^{\prime\prime}&=&\br_{2}^{\prime\prime}-\br_{1}^{\prime\prime} .
\eea
If we assume that the source is quasi-homogeneous and has 
a normalized spectrum that is independent of position, i.e.
\be
{\mathcal W}_{Q}(\bR^{\prime\prime}, \br^{\prime\prime}, \om)=
v_{coh} s_{0}(\om)I_{Q}\left(\bR^{\prime\prime}\right)
\delta(\br^{\prime\prime})
\ee
where $I_{Q}\left(\bR^{\prime\prime}\right)$ is the
intensity distribution, $v_{coh}$ is the coherence volume
(which is formally infinite) and  $s_{0}(\om)$
is the normalized spectrum
then eq.(\ref{george}) can be re-written as follows
\bea
{\mathcal W}(R\bs,\bb,\om)&\approx& \frac{\exp(ik\bs\cdot\bb)}{R^{2}}
v_{coh} s_{0}(\om) \nn \\
&&\times
\int d^{3}R^{\prime\prime}  I_{Q}\left(\bR^{\prime\prime}\right)
\exp\left(\frac{-ik\bb_{\perp}\cdot \bR^{\prime\prime}}{R}\right)
\eea

If we assume that the source is a spherically symmetric Gaussian
function, viz.,
\be
I_{Q}\left(\bR^{\prime\prime}\right)=
I_{0}\exp\left(-\frac{R^{\prime\prime 2}}{2 a^{2}}\right),
\ee
where $a$ can be considered to be the effective radius of the
source,  then we obtain the following expression 
for the cross-spectral density function of the field
in the far zone of the source:
\be
{\mathcal W}(R\bs,\bb,\om)\approx \frac{v_{coh}V_{0}I_{0}}{R^{2}}
s_{0}(\om)\exp(ik\bs\cdot\bb)\exp\left(-\frac{a^{2}k^{2}b^{2}_{\perp}}
{2R^{2}}\right)
\ee
where $V_{0}=(2\pi)^{3/2}a^{3}$ is the volume of the source.
We will treat the baffle shown in fig.2, which represents the
blocking effect of the dust torus, with a purely geometric 
approximation.  Thus the cross-spectral density function
for the field incident upon the scatterer is given by
\be
{\mathcal W}_{0}(R\bs,\bb,\om) = 
\left\{\begin{array}{ll} 
\frac{\D v_{coh}V_{0}I_{0}}{\D R^{2}}
s_{0}(\om)\exp(ik\bs\cdot\bb)\exp\left(-\frac{\D a^{2}k^{2}b^{2}_{\perp}}
{\D 2R^{2}}\right), & {\rm lit\,region} \\
&\\
0, & {\rm shadow\,region} . \end{array} \right.
\label{csdinc}
\ee

\subsection{The Spectral Scattering Kernel}
We will now introduce a model to describe the correlation properties
of the scattering medium.  Let us assume that we can approximate the
cross spectral density of the scatterer by the following formula:
\be
W_{\eta}(\brp_{1},\brp_{2},\om) \approx 
\rho\left(R\right)
\tg(\bb,\bs,\om) ,
\ee
where
\bea
\bR&\equiv&R\bs=\frac{\brp_{1}+\brp_{2}}{2}\nn \\
\bb&=&\brp_{2}-\brp_{1}.
\eea
This model is somewhat analogous to the well-know
quasi-homogeneous model, with the added wrinkle that
the coherence properties can be dependent on orientation
(as characterized by the unit vector $\bs$).  What I
have in mind specifically is that the coherence length
in the radial direction (i.e. along $\bs$) is different
from the coherence length in the transverse direction
(i.e. perpendicular to $\bs$).

Therefore the spectrum is
\be
S^{(\infty)}(r\bu,\om)=\frac{1}{r^{2}}\int\int
d^{3}R d^{3}b \int d\omp \rho(R)
\tg(\bb,\bs,\om) 
{\mathcal W}_{0}(\bR,\bb,\omp)
\exp(ik\bu\cdot\bb) .
\label{specone}
\ee
On substituting from eq.(\ref{csdinc}) into eq.(\ref{specone})
we obtain the following expression for the spectrum of the
scattered radiation
\be
S^{(\infty)}(r\bu,\om)=\frac{1}{r^{2}}d\omp 
\kappa(\bu,\om,\omp) s_{0}(\omp)
\label{spectwo}
\ee
where the {\em spectral scattering kernel} $\kappa(\bu,\om\omp)$
is given by the formula
\be
\kappa(\bu,\om,\omp)\approx
\kappa_{0}\int_{\Omega_{0}}d^{2}\Omega_{\bs}\int d^{3}b \,
\tilde{g}(\bb,\bs,\om-\omp)\exp\left(-\frac{k^{\prime 2}a^{2} b^{2}_{\perp}}
{2R_{0}^{2}}\right) \exp \left[i\left(k^{\prime}\bs-k\bu \right)\cdot\bb\right]
\label{kern}
\ee
where $k^{\prime}=\omp/c$ and 
\be
\kappa_{0}=v_{coh}V_{0}I_{0}\int_{R_{0}}^{\infty}dR\rho(R) .
\ee
In these formulas, $R_{0}$ represents the distance from
the source to the baffle and $\Omega_{0}$ is the solid angle 
of the lit region (see fig.2).

\subsection{The Scattering Medium}
We shall assume that the scattering medium is anisotropic,
in the sense that the correlation length in the radial
direction (i.e. along the vector $\bs$) is different 
from the correlation in the transverse
direction (i.e. perpendicular to $\bs$).  For
simplicity, we will assume a Gaussian model, i.e.
\be
g(\bb,\tau)=\exp\left[-\frac{1}{2}\left(
\frac{b^{2}_{\perp}}{\sigma^{2}_{\perp}}+
\frac{b^{2}_{\parallel}}{\sigma^{2}_{\parallel}}+
\frac{\tau^{2} c^{2}}{\sigma^{2}_{\tau}}
\right)\right] ,
\ee
where $\bb_{\perp}=(\bs.\bb)\bs$
and $\bb_{\parallel}=\bb-\bb_{\perp}$.  
Substituting this formula into eq.(\ref{kern}), we
obtain the following expression for the scattering kernel
\bea
\kappa(\bu,\om,\omp)&\approx&
\kappa_{0}\frac{\sigma_{\tau}}{c\sqrt{2\pi}}
\exp\left[-\frac{1}{2}\frac{\sigma^{2}_{\tau}}{c^{2}}
\left(\om-\omp\right)^{2}\right] \nn \\
&&\times
\int_{\Omega_{0}}d^{2}\Omega_{\bs}\int d^{3}b \,
\exp\left[-\frac{1}{2}
\left(\frac{b_{\perp}^{2}}{\sigma^{\prime 2}_{\perp}}+
\frac{b_{\parallel}^{2}}{\sigma^{2}_{\parallel}}
\right)\right]
\exp \left(i\beff\cdot\bb\right) \nn \\
\\
\label{kerntwo}
\eea
where
\be
\beff=k^{\prime}\bs-k\bu
\ee
and
\be
\frac{1}{\sigma^{\prime 2}_{\perp}}=\frac{1}{\sigma^{2}_{\perp}}+
\frac{a^{2}k^{\prime 2}}{R^{2}}\approx \frac{1}{\sigma^{2}_{\perp}}.
\ee
The approximation made in the last formula is valid provided that the
traverse coherence area of the incident light is much larger than that
of the scattering medium.  Given that the propagation distance $R$ is
of the order on tens or even hundreds of parsecs, this is not an
unreasonable assumption.
Performing the integration over $\bb$ we obtain the following
formula for $\kappa$:
\bea
\kappa(\bu,\om,\omp)&\approx&
\kappa_{0}
\frac{2\pi\sigma_{\tau}\sigma^{2}_{\perp}\sigma_{\parallel}}{c}
\exp\left[-\frac{1}{2}\frac{\sigma^{2}_{\tau}}{c^{2}}
\left(\om-\omp\right)^{2}\right]\nn \\
&&\times\int_{\Omega_{0}}d^{2}\Omega_{\bs}\,
\exp\left(-\frac{1}{2}
\left[
\sigma^{2}_{\perp}k^{2}u_{\perp}^{2}+
\sigma^{2}_{\parallel}(k^{\prime}-ku_{\parallel})^{2}
\right]\right)\nn \\
&=&\kappa_{0}
\frac{(2\pi)^{2} \sigma_{\tau}\sigma^{2}_{\perp}\sigma_{\parallel}}{c}
\exp\left[-\frac{1}{2}\frac{\sigma^{2}_{\tau}}{c^{2}}
\left(\om-\omp\right)^{2}\right]\nn\\
&&\times\int^{1}_{\cos\theta_{0}}dx\,
\exp\left(-\frac{1}{2}
\left[
\sigma^{ 
2}_{\perp}k^{2}(1-x^{2}\cos^{2}\theta^{\prime})\right.\right. \nn\\
&&+\left.\left.
\sigma^{2}_{\parallel}(k^{\prime}-kx\cos\theta^{\prime})^{2}
\right]\right) ,
\label{kernthree}
\eea
where $\theta^{\prime}$ is the scattering angle, as shown
in fig.2.
The integral in eq.(\ref{kernthree}) can be evaluated approximately
using the following formula:
\bea
\int^{1}_{\cos\theta_{0}}f(x)dx&\approx&
(1-\cos\theta_{0})f\left(\frac{1+\cos\theta_{0}}{2}\right) \nn \\
&=&\frac{\Omega_{0}}{2\pi}
f\left(1-\frac{\Omega_{0}}{4\pi}\right).
\eea
Then we obtain the following
\be
\kappa(\bu,\om,\omp)\approx
\kappa_{0}
\frac{2\pi \sigma_{\tau}\sigma^{2}_{\perp}\sigma_{\parallel}\Omega_{0}}{c}
\exp\left(-\frac{1}{2}
\left[
\alpha^{\prime}\omega^{\prime 2}-2\beta\om\omp+\alpha\om^{2}
\right]\right)
\label{kernfour}
\ee
where
\bea
\alpha&=&\frac{1}{c^{2}}\left[
\sigma_{\tau}^{2}+\sigma_{\perp}^{2}+
q^{2}(\sigma_{\parallel}^{2}-\sigma^{2}_{\perp})\right]\nn \\
\beta&=&\frac{1}{c^{2}}\left[
\sigma_{\tau}^{2}+ q\sigma_{\parallel}^{2}\right] \nn \\
\alpha^{\prime}&=&\frac{1}{c^{2}}\left[
\sigma_{\tau}^{2}+ \sigma_{\parallel}^{2}\right] \nn \\
q&=&\left(1-\frac{\Omega_{0}}{4\pi}\right)\cos\theta^{\prime}
\label{alphabeta}
\eea
\subsection{The z-number of the scattered radiation}
The scattering kernel given in eq.(\ref{kernfour}) is equivalent
to one we have already investigated in a previous paper [see
ref.\cite{S:7}, eq.(12)].  In that paper, we showed that such a
spectral scattering kernel causes Doppler-like spectral shifts
with $z$-numbers given by the formula:
\be
z=\frac{\alpha}{|\beta|}-1 .
\ee
Substituting the expressions for $\alpha$ and $\beta$ from
eq.(\ref{alphabeta}), we obtain the following result:

\be
z=\frac{1-q}{\sigma^{2}_{\tau}+\sigma^{2}_{\parallel}q}
\left[\sigma^{2}_{\perp}+q(\sigma^{
2}_{\perp}-\sigma^{2}_{\parallel})\right]
\ee

Let us assume a `white noise limit' for the scattering
medium: i.e. its time fluctuations are effectively 
uncorrelated:
$\sigma_{\tau}\rightarrow 0$.  Then the formula for the spectral
shift of the scattered radiation is:
\be
z(\varepsilon, q)=\frac{\varepsilon}{q}+(1-\varepsilon)q-1 ,
\label{monty}
\ee
where
\bea
\varepsilon&=&\left(\frac{\sigma_{\perp}}{\sigma_{\parallel}}\right)^{2}\\
q&=&\left(1-\frac{\Omega_{0}}{4\pi}\right)\cos\theta^{\prime}.
\eea
Equation (\ref{monty}), which is displayed graphically
in Fig.3, is our final result.  It demonstrates that, for the 
specific model of active galactic nucleus we are considering, the
spectral shift observed will be dependent on the viewing angle
$\theta^{\prime}$, the solid angle of the radiation cone $\Omega_{0}$
and the statistical anisotropy of the scatterer, as represented by the
parameter $\varepsilon$.  When $\varepsilon=1$, the scattering medium
is statistically isotropic: the coherence lengths are the same in all 
directions.  In the limit $\varepsilon\rightarrow 0$ the longitudinal
coherence length is much greater than the transverse coherence 
length, i.e. one has a cigar shaped coherence volume.  In the 
limit $\varepsilon\rightarrow \infty $, one has the opposite case: the longitudinal
coherence length is much smaller than the transverse coherence 
length and the coherence volume is like a pancake.  Note that
$z(\varepsilon, q)>0$ provided that $\varepsilon>0.5$, so that
one will have a preference redshifts over a large area of the
$(\varepsilon, q)$ plane.

\section{Discussion and Conclusions}

One can easily spot potential problems with the theory presented here.
For example, we have been deliberately vague about the
underlying physical nature of the scattering medium,
other than specifying its anisotropic coherence properties.
The scatterer, which is assumed to have a `white noise' 
power spectrum (implying that its fluctuations are very
energetic), is situated further away from the central engine
than the line emitting clouds (which cannot be too hot, otherwise
they would have completely ionized, making line radiation 
impossible).  Although our results apply for a broad spectral
range, we have not considered the shifts of absorption lines or
of the 21cm radio line.  We have ignored the unscattered
radiation and the efficiency of the scattering process
(although it is possible a stimulated version of
the spontaneous scattering process considered here
is applicable).  Also the issue of spectral linewidths has
not been addressed.  However, our results do
demonstrate the plausibility of the existence of
extra galactic objects with discordant redshifts
due to well established physical phenomena.

It should be emphasized that our theory is based
entirely on established principles of optics: no
hypothetical new physics was introduced.  Furthermore
certain aspects of the underlying phenomenon, the
Wolf effect, have been extensively tested in the
laboratory.  The formula (\ref{monty}) gives the
relationship between the $z$-number of the observed
radiation and certain variables dependent on the
object, namely the angle of vision, the degree of
collimation due to the dust torus, and the degree
of anisotropy of the scatterer.  As it may be possible
to measure some of these quantities for certain objects,
it therefore may be possible to test these predictions.

\section*{Acknowledgments}
This paper, like the others in this special issue, is presented in 
honor of Prof. Emil Wolf on the occasion of his 75th birthday.  It
was my great good fortune to be Prof. Wolf's research student between 1987 and 
1992, and I learned a tremendous amount from him.  His 
enormous knowledge of all aspects of optical physics and
his care and attentiveness as a mentor are outstanding. 
A lot of the ideas presented in this paper originate 
from his work.

\newpage
\section*{Figure Captions}
Fig.1: A schematic illustration of the model AGN under consideration.
\\
\\
Fig. 2: A simplified version of Fig.1, for mathematical analysis,
showing the symbols used.
\\
\\
Fig.3:  The relationship between $z$ number and the collimation
factor $q$ for different anisotropy factors $\varepsilon$, 
illustrating the result eq.(\ref{monty}).

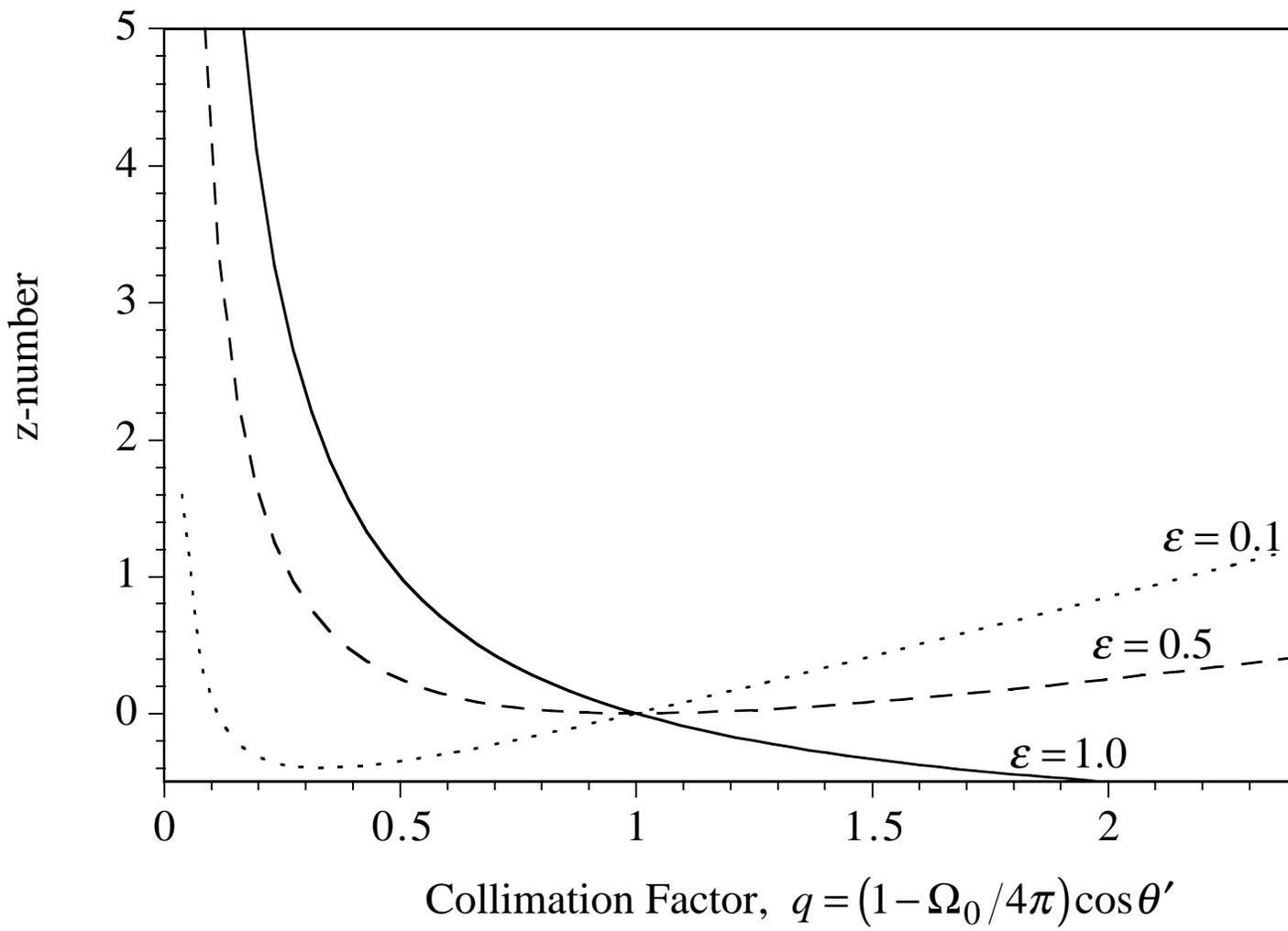
\end{document}